\documentclass{optica-article}

\journal{opticajournal} % for journals or Optica Open

\articletype{Research Article}

\usepackage{lineno}
\usepackage{glossaries}
\glsdisablehyper
\newacronym{rpi}{RPI}{randomized probe imaging}
\newacronym{rzp}{RZP}{randomized zone plate}
\newacronym{cdi}{CDI}{coherent diffraction imaging}
\newacronym{fth}{FTH}{Fourier transform holography}
\newacronym{fel}{FEL}{free electron laser}
\newacronym{ssnr}{SSNR}{spectral signal-to-noise ratio}
\newacronym{ilh}{ILH}{in-line holography}
\newacronym{sbp}{SBP}{space-bandwidth product}
\newacronym{ssp}{SSP}{single-shot ptychography}

\usepackage{xr}

\usepackage{siunitx}
\DeclareSIUnit\bar{bar}

\usepackage{xurl}

\begin{document}

\title{Single-shot imaging with randomized structured illumination at a free electron laser}

\author{Abraham L. Levitan, \authormark{1,2*} Kahraman Keskinbora,\authormark{1} Matteo Pancaldi,\authormark{3,4} Dieter W. Engel, \authormark{5} Emanuele Pedersoli,\authormark{3} Flavio Capotondi,\authormark{3} and Riccardo Comin\authormark{1}}

\address{\authormark{1}Massachusetts Institute of Technology, 77 Mass Avenue, Cambridge, MA 02139, USA\\
\authormark{2}Paul Scherrer Institute, Forschungsstrasse 111, 5232 Villigen PSI,Switzerland\\
\authormark{3}Elettra-Sincrotrone Trieste, Strada Statale 14 km 163,5 in Area Science Park, I-34012 Basovizza, Trieste, Italy\\
\authormark{4}Department of Molecular Sciences and Nanosystems, Ca' Foscari University of Venice, I-30172 Venezia Mestre, Italy\\
\authormark{5}Max-Born-Institut für Nichtlineare Optik und Kurzzeitspektroskopie, Max-Born-Straße 2A, 12489 Berlin, Germany
}

\email{\authormark{*}alevitan@mit.edu} %% email address is required; see note below about the corresponding author designation

% use {asbstract*} to suppress the copyright line. Copyright information will be added in production

\begin{abstract*} 
Stroboscopic nanoscale imaging with free electron laser light is revolutionizing our understanding of fast dynamics in heterogeneous systems. The short wavelength of X-ray and extreme ultraviolet radiation makes it possible to achieve nanoscale resolution, while resonance with atomic transitions gives access to electronic and magnetic degrees of freedom. Here, we report on our implementation of a recently developed imaging method, randomized probe imaging, at a free electron laser. The advantage of randomized probe imaging over existing methods is its compatibility both with extended and strongly scattering samples. Our implementation delivers robust single-shot reconstructions at up to a full-pitch resolution of $\SI{400}{\nano\meter}$ over a field of view with a $\SI{40}{\micro\meter}$ diameter. We also demonstrate single-shot imaging of magnetic domain structures using circular dichroism at resonance, paving the way to future time-resolved studies of magnetic dynamics, shock physics, and the dynamics of collective electronic phases.
\end{abstract*}

%%%%%%%%%%%%%%%%%%%%%%%%%%  body  %%%%%%%%%%%%%%%%%%%%%%%%%%
\glsresetall %This resets the expansion of acronyms after the abstract

\section{Introduction}

Following the 2006 experiment performed by Chapman et al \cite{chapman2006}, X-ray \gls{cdi} \cite{fienup1987, miao1999a} has been used at \glspl{fel} to image an astonishing variety of phenomena. A program of research that began by imaging simple test objects \cite{chapman2006, barty2008} was soon studying the morphology of metallic nanoparticles \cite{takahashi2013a, clark2015a, ihm2019}, soot \cite{bogan2010,loh2012}, helium nanodroplets \cite{jones2016}, battery electrolytes \cite{suzuki2022}, viruses \cite{seibert2011, gorkhover2018}, organelles \cite{hantke2014,takayama2015,pan2022}, and even whole bacterial cells \cite{kimura2014, vanderschot2015, kobayashi2021}.

The introduction of other, related quantitative phase imaging methods at \glspl{fel} broadened the possibilities further. For example, taking \gls{ilh} \cite{garcia-sucerquia2006, gorniak2011} data---which can then be analyzed with various algorithms \cite{abbey2008, williams2010, vassholz2021, hagemann2021}---enables experimenters to work with extended (i.e. non-isolated) samples. \gls{ilh} has therefore been very useful for visualizing the mechanical properties of matter undergoing rapid changes, e.g. propagating shock waves \cite{sandberg2014, schropp2015, nagler2016, seiboth2018}, cavitation bubbles \cite{vassholz2021}, liquid microjets \cite{hagemann2021}, and exploding capillaries \cite{vagovic2019}.

Going in a different direction, \gls{fth} \cite{mcnulty1992, eisebitt2004} and its derivatives \cite{marchesini2008, guizar-sicairos2008b, martin2014} introduce extra sample masking and preparation steps but produce data which can be reconstructed more reliably, and with less computational cost, than either \gls{cdi} or \gls{ilh} data can. This is especially true when the sample is a strong scatterer or contains densely packed features, both of which would destabilize an \gls{ilh} reconstruction. For this reason, \gls{fth} has become popular for studies using extreme ultraviolet and soft X-ray light, where the strong resonant coupling to electronic degrees of freedom has enabled studies of magnetism in thin films \cite{muller2013, willems2017}, including picosecond-scale dynamics \cite{vonkorffschmising2014}.

For quantitative phase imaging at an \gls{fel} today, one can choose to either image an extended sample or a strongly scattering sample, but not both at once. As a result, for phase imaging there is no standard \gls{fel}-based method with the same robustness and simplicity as \gls{fel}-based full-field absorption contrast imaging \cite{egawa2024}. This is the motivation behind our \gls{fel}-based implementation of \gls{rpi}, a quantitative phase imaging method that works reliably with extended, strongly scattering samples \cite{levitan2020}. 

The ability to work in the same regime of quality and resolution as existing methods, which we demonstrate with a comprehensive literature search, is the main distinction between \gls{rpi} and the exciting recent report of an \gls{fel} based implementation of \gls{ssp} \cite{sidorenko2016, kharitonov2022}. Furthermore, \gls{rpi} is robust, with a well-posed inverse problem that rapidly converges even on complicated samples. The experiment itself is simple and can be done at nearly any coherent X-ray scattering instrument which supports ptychography. No optics need to be placed in the Fresnel regime of the sample, creating space for special sample environments and optical photoexcitation. Finally, \gls{rpi} is scalable: the inverse problem remains well-posed even if a large field of view is imaged in high resolution \cite{levitan2020}. 

In the experiment reported here we show that these advantages carry over from the synchrotron to the \gls{fel} context, enabling single-shot imaging of both structural and magnetic samples. On the structural front, we reconstruct complex-valued images of Siemens star test samples, enabling us to assess the quality of our \gls{rpi} results in detail and explore the prospects for future implementations. In addition, we report images of magnetic domain structures in a thin Pt/Co/Ta multilayer sample, demonstrating the basic applicability of the method to challenging samples with weak contrast. In sum, \gls{rpi}'s unique combination of advantages makes it a compelling alternative to \gls{cdi}, \gls{ilh}, and \gls{fth} for a variety of experiments across the wavelength range used at \glspl{fel}.

\section{Methods}

\begin{figure}
\centerline{\includegraphics[width=180mm]{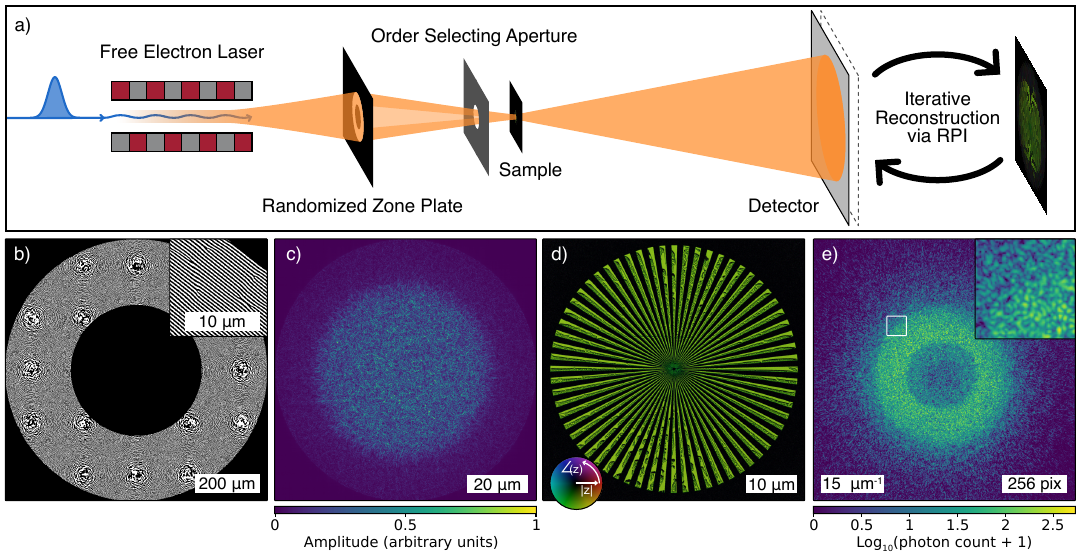}}
\caption[Overview of the \gls{fel} \gls{rpi} experiment.]{\textbf{Experiment Overview.} (a) A diagram of the experiment's geometry and workflow. (b) The design file for the optic used in the experiment, sampled in 1 um steps. Inset: a section of the outer region, showing typical phase defects. (c) The probe function at the sample plane, as retrieved via ptychography. (d) The sample's complex-valued object function, as retrieved via ptychography. (e) A typical single-shot diffraction pattern, with the inset showing the detailed speckle structure.}\label{fig:overview}
\end{figure}

The workflow that we established for single-shot \gls{fel}-based \gls{rpi} experiments is shown in Figure \ref{fig:overview}a. A single \gls{fel} pulse passes through a \gls{rzp} \cite{levitan2020} (Figure \ref{fig:overview}b), which projects a speckled structured illumination function (Figure \ref{fig:overview}c) onto the sample. The upstream optics at the beamline are set to overfill the \gls{rzp} by several times, helping to mitigate the shot-to-shot wavefront instability of the \gls{fel}. Further information on the design and fabrication of this optic is found in supplemental section S1. In this work, we first used a roughly $\SI{60}{\femto\second}$ long pulse of $\SI{20.8}{\nano\meter}$ wavelength light produced by the FERMI \gls{fel} to image a gold Siemens star (Figure \ref{fig:overview}d) on a $\SI{30}{\nano\meter}$ thick silicon nitride window. After passing through the sample, the light propagates to the far field where the diffracted intensity is recorded on a detector. A typical raw single-shot pattern, consisting of approximately $1.2\times10^7$ measured photons and arising from an estimated incident pulse energy at the sample plane of $\SI{1.4}{\nano\joule}$, is shown in Figure \ref{fig:overview}e. Further information about the detailed experimental setup is found in supplemental section S2.

Detailed knowledge of the illumination wavefield is obtained empirically by reconstructing a standard transmission ptychography dataset \cite{pfeiffer2018} collected with one shot per exposure on a static sample. Here, this ptychography reconstruction was performed with the cdtools package \cite{levitanCDTools2024} using an automatic differentiation-based reconstruction algorithm which included 3 incoherently mixing probe modes \cite{thibault2013}, probe position refinement with a gradient-based algorithm \cite{tripathi2014}, a shot-to-shot total intensity factor, a frame-independent detector background model \cite{marchesini2013}, and a super-resolution approach inspired by \cite{maiden2011}. The exact details of the reconstruction algorithm can be found in the publicly available replication data and code for this work \cite{levitan2026}. Super-resolution was used to avoid aliasing artifacts because the scattered intensity remained high at the detector edges.

Using the Fourier ring correlation, we determined that the ptychography reconstruction of the Siemens star target was reliable to the super-resolution pixel pitch of  $\SI{73}{\nano\meter}$, a full-pitch resolution of $\SI{146}{\nano\meter}$. We also established the reliability of the recovered probes, finding a normalized amplitude-minimized partially coherent mean squared error \cite{levitan2022} of $9.1\%$ between probes recovered from each half of the dataset. Various additional assessments of the recovered ptychography quality are shown in supplementary Figure S4.

Following this calibration step, the probe function recovered from ptychography is used in conjunction with the \gls{rpi} algorithm \cite{levitan2020} to reconstruct images of the illuminated region of the Siemens star test object on a shot-to-shot basis. This algorithm, described in full in \cite{levitan2020}, uses a band-limiting constraint on the object together with knowledge of the illumination function to stabilize the phase retrieval problem. The single-shot \gls{rpi} reconstructions were performed with the cdtools package \cite{levitanCDTools2024}, using $650\times650$ pixel objects with a pixel size of $\SI{160}{\nano\meter}$. As discussed in \cite{levitan2020}, the resolution of the single-shot \gls{rpi} reconstructions is roughly limited by the length scale of the speckles in the structured illumination, which is ultimately tied to the outer zone width of the \gls{rzp}. In this work, the optic's outer zone width was $\SI{200}{\nano\meter}$, and as we demonstrate next, the design full-pitch resolution of $\SI{400}{\nano\meter}$ was achieved.

\section{Results}

\begin{figure}
\centerline{\includegraphics[width=180mm]{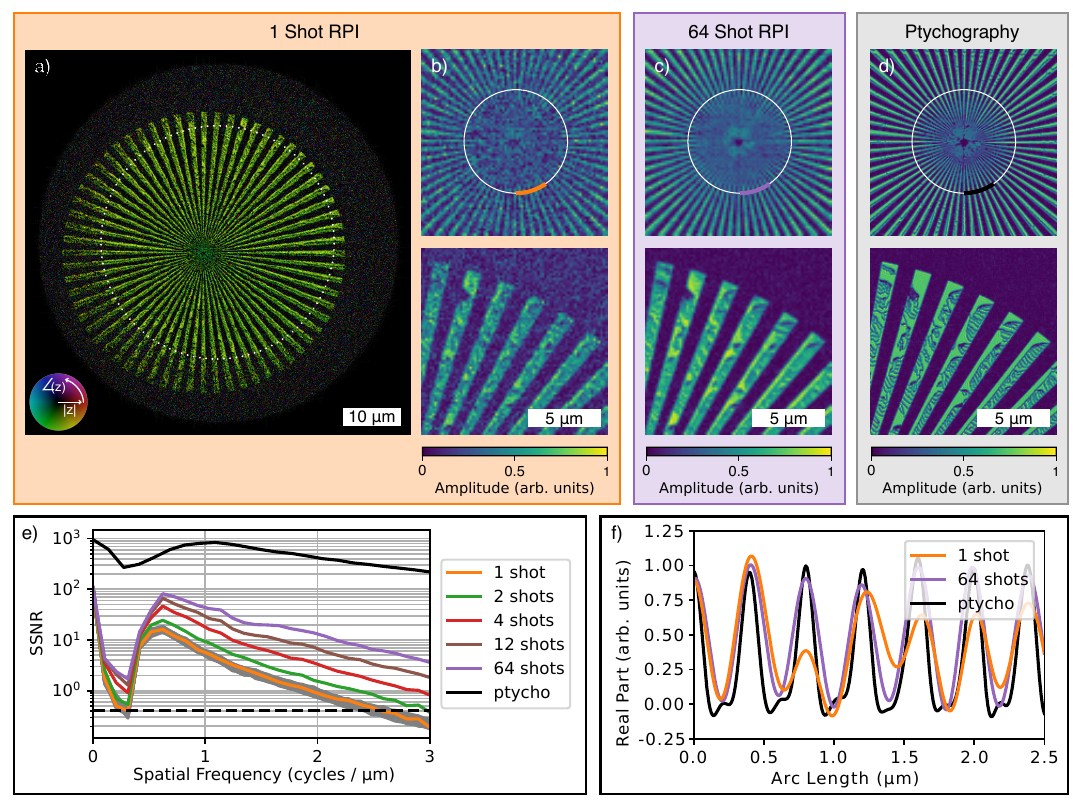}}
\caption[A single-shot \gls{rpi} reconstruction]{\textbf{RPI Reconstruction}. (a) The complex-valued result of a typical single-shot reconstruction. The approximate illuminated region is outlined with a white dashed line. (b) Top, a closeup of the amplitude within the central region. The white circle intersects the spokes at a $\SI{400}{\nano\meter}$ pitch. Bottom, an outer region of the field of view. (c) The same regions, extracted from the mean of 64 single-shot images. (d) The same regions, extracted from the calibration ptychography reconstruction. (e) The \gls{ssnr} of the single-shot reconstruction, as well as various summed images and the ptychography reconstruction. The grey lines are the results from all 128 single-shot reconstructions, and the black dashed line indicates the half-bit threshsold (f) Line cuts extracted along the paths defined in b,c, and e.}\label{fig:rpi}
\end{figure}

A typical single-shot \gls{rpi} reconstruction from this setup is shown in Figures \ref{fig:rpi}a and \ref{fig:rpi}b. Despite the presence of noise, the good subjective visual quality is immediately apparent, as is the lack of obvious artifacts. Furthermore, the reconstructions are remarkably consistent. The image in Figure \ref{fig:rpi}a was inverted from the first of a series of 128 consecutive exposures. The reconstruction from each of the 128 exposures attained a similar quality level, and all 128 images are presented in supplementary figures S6 and S7.

We evaluated the quality of these images using the \gls{ssnr} \cite{penczek2010}, shown in Figure \ref{fig:rpi}e. This was estimated from a Fourier cross resolution curve \cite{penczek2010} calculated between the \gls{rpi} results and an appropriately downsampled ptychography reconstruction of the same sample using the methodology described in supplementary note S4. The resulting curve has a characteristic shape with a dip at intermediate spatial frequencies, due to the lack of these frequencies in the Siemens star test object. To estimate the resolution attained, we compared the \gls{ssnr} curve with a half-bit threshold of $0.4142$ \cite{vanheel2005}, commonly used to assess the resolution of lensless imaging data \cite{vila-comamala2011, guizar-sicairos2012, donnelly2016, shapiro2020}. By this metric, the vast majority of the 128 reconstructions surpassed the design full-pitch resolution of $\SI{400}{\nano\meter}$, corresponding to the optic's $\SI{200}{\nano\meter}$ outer zone width. The worst \gls{ssnr}-determined resolution among the full set of 128 reconstructions was $\SI{420}{\nano\meter}$.

Each single-shot diffraction pattern contained an average of $1.2\times10^7$ detected photons, equivalent to $490$ photons per $\SI{200}{\nano\meter}$ resolution element within the illuminated region. To understand how the quality of \gls{rpi} reconstructions would improve as the fluence increases, we averaged the results of subsequent single-shot reconstructions after aligning the images to correct for the roughly $\SI{1}{\micro\meter}$ positioning jitter. An example summed image, generated from 64 reconstructions and $3.1\times10^4$ photons per resolution element, is shown in Figure \ref{fig:rpi}c and compared to a corresponding ptychography reconstruction shown in Figure \ref{fig:rpi}d. 

The \gls{ssnr} improved as we included more exposures in the summed image, with the quality saturating around 64 exposures. Quantitatively, the normalized mean squared error \cite{fienup1997} calculated between summed \gls{rpi} images and downsampled ptychography images saturated just under $5\%$. This implies that the quality was ultimately limited by other noise sources, such as the presence of unaccounted-for high spatial frequency components in the object function or potentially insufficient shot-to-shot stability of the illumination wavefield. For a sum of 64 exposures the images were resolved pixel-by-pixel, with an \gls{ssnr} at the pixel pitch of roughly 3 as shown in Figure \ref{fig:rpi}e.

For a more direct confirmation that the design resolution was achieved in single-shot images, in Figure \ref{fig:rpi}f we used the methodology described in supplementary note S5 to plot a cut from the image along an arc which intersects the spokes of the Siemens star at a $\SI{400}{\nano\meter}$ pitch. The periodic pattern of spokes is clearly resolved, supporting our claim of a $\SI{400}{\nano\meter}$ full-pitch resolution.

Following the imaging experiments with resolution test targets, we assessed the viability of \gls{fel} based \gls{rpi} for resonant imaging of magnetic thin films. To do this, we collected similar ptychography and \gls{rpi} datasets from a magnetic thin film using circular left- and right-hand illumination at the Co M3 edge. The multilayer thin film was prepared with a nominal structure of Ta($\SI{3}{\nano\meter}$)/[Pt($\SI{3}{\nano\meter}$)/Co($\SI{1}{\nano\meter}$)/Ta($\SI{1.9}{\nano\meter}$)]x10/Pt($\SI{2}{\nano\meter}$), deposited on a $\SI{30}{\nano\meter}$ thick silicon-rich silicon nitride membrane using direct current and radio frequency magnetron sputtering. During deposition, the argon pressure was adjusted to $4.6\times10^{-3}\si{\milli\bar}$. This preparation hosts natural magnetic domains at remanence, covering a variety of length scales from several hundred $\si{\nano\meter}$ to several $\si{\micro\meter}$.

\begin{figure}
\centerline{\includegraphics[width=88mm]{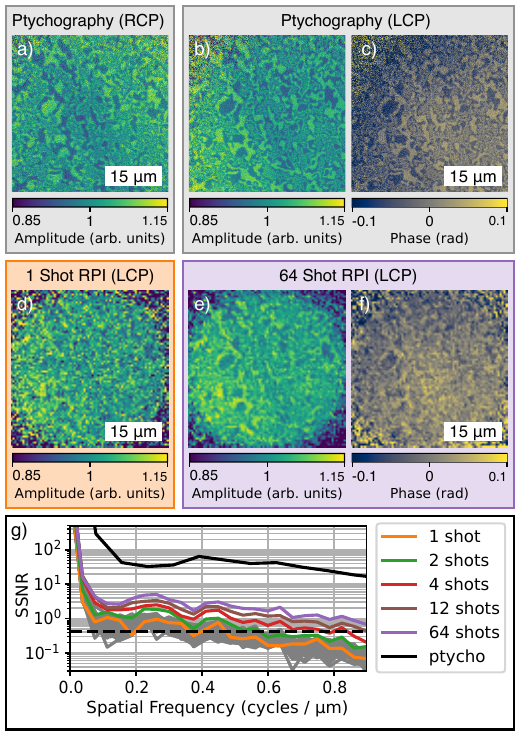}}
\caption[Ptychography and \gls{rpi} reconstructions of a magnetic thin-film sample]{\textbf{Imaging of a magnetic film}. (a) The amplitude of a ptychography reconstruction from a Pt/Co/Ta multilayer under right-hand circularly polarized illumination. The natural magnetic domain structure is visible. (b) The amplitude and (c) phase of a ptychography reconstruction of the same region under left-hand circular illumination. (d) Single-shot \gls{rpi} from the same region. (e) The amplitude and (f) phase of an average of 64 \gls{rpi} reconstructions. (g) The \gls{ssnr} of the ptychography and \gls{rpi} reconstructions, with varying numbers of summed images. The grey lines are the results from all 64 single-shot reconstructions, and the black dashed line indicates the half-bit threshold.}\label{fig:magnetic_rpi}
\end{figure}

Ptychography reconstructions with right-hand and left-hand circularly polarized illumination are shown in Figures \ref{fig:magnetic_rpi}a-\ref{fig:magnetic_rpi}c. The contrast reversal demonstrates the magnetic origin of the imaged structures. The left-hand circular results, which were used to calibrate the \gls{rpi} reconstructions, achieved a resolution of $\SI{390}{\nano\meter}$ as determined by \gls{ssnr}. Various standard assessments of the recovered image quality are shown in supplementary figure S5.

A series of 64 shots of \gls{rpi} data were then captured under left-hand circularly polarized illumination, with a mean of $2.8\times10^6$ photons measured per shot. \gls{rpi} reconstructions were performed with a $200\times200$ pixel object with a pixel size of $\SI{520}{\nano\meter}$. An example single-shot \gls{rpi} reconstruction is shown in Figure \ref{fig:magnetic_rpi}d, and an average of all 64 individual \gls{rpi} reconstructions is shown in Figures \ref{fig:magnetic_rpi}e and \ref{fig:magnetic_rpi}f. Notably, both amplitude and phase contrast arising from the magnetic domains is visible in the averaged \gls{rpi} images. \gls{ssnr} curves calculated via comparison between \gls{rpi} and ptychography results are shown in Figure \ref{fig:magnetic_rpi}g, indicating that the single-shot reconstructions achieved a full-pitch resolution of $\SI{2.22}{\micro\meter}$, and averages of 12 or more images achieved the pixel-limited full-pitch resolution of $\SI{1.04}{\micro\meter}$.

Finally, we estimated the \gls{sbp} of the single-shot reconstructions. \Gls{sbp} is a measure of the number of individually resolved resolution elements contained within the image \cite{lohmann1996}. It is defined as the product of the accessible area in real space and reciprocal space:

\begin{equation}
\text{SBP} = A_\text{real} A_\text{Fourier} = \pi^2 \frac{r^2}{\text{res}^2}.
\end{equation}

Conservatively estimating our field of view as a $\SI{40}{\micro\meter}$ diameter focal spot, we find a \gls{sbp} of $2.5\times10^4$ for the single-shot structural images,  $3.6\times10^3$ for the 64 shot magnetic images, and $8.0\times10^2$ for the single-shot magnetic images. These are equivalent to a perfectly resolved $160\times160$ pixel image, $60\times60$ pixel image, and $28\times28$ pixel image, respectively.

\section{Discussion}

We have implemented \gls{rpi} at an \gls{fel}, showing that \gls{rpi} accommodates noisy conditions gracefully and that it is possible to collect high-quality images with sufficiently high flux. The reliability of the method was demonstrated by the remarkably consistent quality achieved across reconstructions from 128 consecutive exposures. In addition, we demonstrated that the method is sufficiently sensitive to image the X-ray circular magnetic dichroic contrast in magnetic domain structures, albeit at a reduced resolution. Combined with the various other advantages of \gls{rpi}---a simple experimental geometry, applicability to samples with strong phase and amplitude features, and compatibility with extended samples---this already makes \gls{rpi} unusual in the world of \gls{fel}-based quantitative phase imaging methods. As we discuss next, \gls{rpi} is also competitive with existing methods for single-shot quantitative phase imaging at \glspl{fel} when considering metrics for image quality, namely resolution and \gls{sbp}.

To understand how our implementation compares with the state of the art, we performed a comprehensive literature search to find all published quantitative phase images collected at \glspl{fel} using methods compatible with single-shot operation, at any wavelength from extreme ultraviolet to hard X-ray. We identified 102 datasets from 97 papers published before November 12, 2023. We estimated the resolution and \gls{sbp} of the highest-resolution image in each dataset, and organized the images into several groups by imaging method. Our methodology for this literature search is described in supplementary note S5.

\begin{figure}
\centerline{\includegraphics[width=88mm]{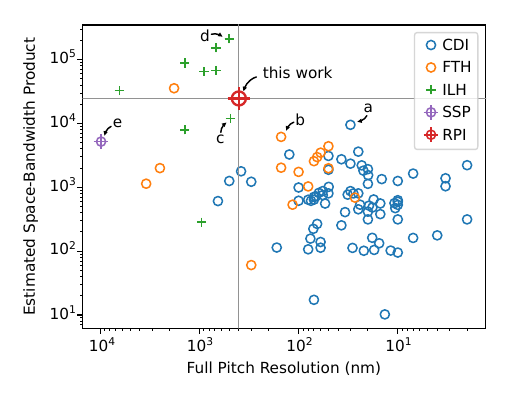}}
\caption{\textbf{Comparison}. A comparison of our single-shot RPI experiment on the gold test structure to all published quantitative phase images collected at \glspl{fel} using single-shot compatible methods as of November 2023. Circular markers (\gls{cdi} and \gls{fth}) indicate methods which are applicable to strong scatterers. Crosses (\gls{ilh}) indicate maskless experiments. Methods which have both properties (\gls{ssp} and \gls{rpi}) have a combined marker. Several relevant experiments are explicitly labeled, and the reader is encouraged to compare our work to these results to understand the qualitative improvement in visual quality afforded by \gls{rpi}. a: \cite{martin2012} (Figure 3); b: \cite{capotondi2013} (Figure 4h); c: \cite{gorniak2011} (Figure 3); d: \cite{vassholz2021} (Figure 2d); and e \cite{kharitonov2022} (Figure 4c).}\label{fig:lit_comparison}
\end{figure}

The results in Figure \ref{fig:lit_comparison} reveal \gls{rpi}'s unique capabilities. Clustered in the upper left region of the plot---with high \gls{sbp} but low resolution---are \gls{ilh} based studies. Like \gls{rpi}, \gls{ilh} works on extended samples, but notably the resolution we achieve is higher than the resolution achieved by any published quantitative phase image from an \gls{fel}-based \gls{ilh} experiment. We note that synchrotron-based \gls{ilh} experiments in the hard X-ray regime---not covered by this literature review---do routinely achieve resolutions in the tens of nanometers.

In the lower right region of the plot lie most \gls{cdi} and \gls{fth} reconstructions. \gls{cdi} and \gls{fth} can achieve astonishingly high resolutions, however, they are restricted to isolated samples. Further, both methods, especially \gls{cdi}, are limited in \gls{sbp}. Therefore, no \gls{fel}-based \gls{cdi} image, and only one \gls{fel}-based \gls{fth} image \cite{pfau2010}, has ever reached a higher \gls{sbp} than that of our recovered images.

The only truly comparable method---both maskless and applicable to strongly scattering samples---that has been used for quantitative phase imaging at an \gls{fel} is \gls{ssp} \cite{kharitonov2022}. The reconstructions from our implementation achieve a far higher resolution and \gls{sbp} than the current state of the art in \gls{fel}-based \gls{ssp}.

Considering the plot as a whole, our \gls{rpi} implementation lies on the Pareto frontier of resolution and \gls{sbp} for \gls{fel}-based experiments. Improvements along both axes are also feasible. The half-pitch resolution is ultimately limited by the outer zone width of the \gls{rzp} used. Zone plates with $\SI{20}{\nano\meter}$ outer zone widths are not uncommon. The \gls{sbp} is ultimately limited by the detector pixel count, photon budget, and/or the source bandwidth. With current detector and source technologies, we expect that a \gls{sbp} in excess of $10^5$ is achievable. 

In summary, \gls{rpi} can be implemented at a seeded \gls{fel} source, and the resulting images are both competitive with state of the art methods and applicable to meaningful samples, such as systems of magnetic domains in extended thin films. We expect that \gls{rpi}'s simplicity, reliability, and compatibility with unmasked samples will make it an attractive option for many stroboscopic imaging experiments that otherwise would have used \gls{fth}, \gls{cdi}, or \gls{ilh}.

\begin{backmatter}
\bmsection{Funding}
This material is based upon work supported by the Department of Energy, Office of Science, Office of Basic Energy Sciences, under Award Number DE-SC0021939. The P6000 GPU used for the reconstructions in this paper was donated by the NVIDIA Corporation. K.K. acknowledges funding support by the German Research Foundation (DFG) under project number 428809035.

\bmsection{Author Contribution Statement}

AL, KK, and RC developed the concept for and organized the experiment. AL and KK designed and KK fabricated the optics. DE grew the magnetic thin film sample. The experiment setup was designed and installed by MP, EP, and FC. AL, KK, MP, EP, and FC participated in data collection. The data analysis, reconstructions, and literature review were performed by AL. All authors contributed to the manuscript.

\bmsection{Acknowledgment}
We thank Dr. Yang Yu from Raith US and MIT.nano for his invaluable help with the VELION ion beam lithography instrument.

\bmsection{Disclosures}
\noindent E: KK now works for Raith Group, which is a supplier of the dedicated IBL system used in this work.

\bmsection{Data availability} 
The raw data, analysis scripts, and literature search data used in this paper have been made available at reference \cite{levitan2026}.

\bmsection{Supplemental document}
See Supplement 1 for supporting content.

\end{backmatter}

\section{References}

%%%%%%%%%% If using BibTeX:
\bibliography{main}

\begin{thebibliography}{10}
\newcommand{\enquote}[1]{``#1''}

\bibitem{chapman2006}
H.~N. Chapman, A.~Barty, M.~J. Bogan, \emph{et~al.}, \enquote{Femtosecond diffractive imaging with a soft-{{X-ray}} free-electron laser,} {\protect\JournalTitle{Nature Physics}} \textbf{2}, 839--843 (2006).

\bibitem{fienup1987}
J.~R. Fienup, \enquote{Reconstruction of a complex-valued object from the modulus of its {{Fourier}} transform using a support constraint,} {\protect\JournalTitle{Journal of the Optical Society of America A}} \textbf{4}, 118 (1987).

\bibitem{miao1999a}
J.~Miao, P.~Charalambous, J.~Kirz, and D.~Sayre, \enquote{Extending the methodology of {{X-ray}} crystallography to allow imaging of micrometre-sized non-crystalline specimens,} {\protect\JournalTitle{Nature}} \textbf{400}, 342--344 (1999).

\bibitem{barty2008}
A.~Barty, S.~Boutet, M.~J. Bogan, \emph{et~al.}, \enquote{Ultrafast single-shot diffraction imaging of nanoscale dynamics,} {\protect\JournalTitle{Nature Photonics}} \textbf{2}, 415--419 (2008).

\bibitem{takahashi2013a}
Y.~Takahashi, A.~Suzuki, N.~Zettsu, \emph{et~al.}, \enquote{Coherent {{Diffraction Imaging Analysis}} of {{Shape-Controlled Nanoparticles}} with {{Focused Hard X-ray Free-Electron Laser Pulses}},} {\protect\JournalTitle{Nano Letters}} \textbf{13}, 6028--6032 (2013).

\bibitem{clark2015a}
J.~N. Clark, L.~Beitra, G.~Xiong, \emph{et~al.}, \enquote{Imaging transient melting of a nanocrystal using an {{X-ray}} laser,} {\protect\JournalTitle{Proceedings of the National Academy of Sciences}} \textbf{112}, 7444--7448 (2015).

\bibitem{ihm2019}
Y.~Ihm, D.~H. Cho, D.~Sung, \emph{et~al.}, \enquote{Direct observation of picosecond melting and disintegration of metallic nanoparticles,} {\protect\JournalTitle{Nature Communications}} \textbf{10}, 2411 (2019).

\bibitem{bogan2010}
M.~J. Bogan, S.~Boutet, H.~N. Chapman, \emph{et~al.}, \enquote{Aerosol {{Imaging}} with a {{Soft X-Ray Free Electron Laser}},} {\protect\JournalTitle{Aerosol Science and Technology}} \textbf{44}, i--vi (2010).

\bibitem{loh2012}
N.~D. Loh, C.~Y. Hampton, A.~V. Martin, \emph{et~al.}, \enquote{Fractal morphology, imaging and mass spectrometry of single aerosol particles in flight,} {\protect\JournalTitle{Nature}} \textbf{486}, 513--517 (2012).

\bibitem{jones2016}
C.~F. Jones, C.~Bernando, R.~M.~P. Tanyag, \emph{et~al.}, \enquote{Coupled motion of {{Xe}} clusters and quantum vortices in {{He}} nanodroplets,} {\protect\JournalTitle{Physical Review B}} \textbf{93}, 180510 (2016).

\bibitem{suzuki2022}
A.~Suzuki, H.~Tanaka, H.~Yamashige, \emph{et~al.}, \enquote{Femtosecond {{X-ray Laser Reveals Intact Sea}}--{{Island Structures}} of {{Metastable Solid-State Electrolytes}} for {{Batteries}},} {\protect\JournalTitle{Nano Letters}} \textbf{22}, 4603--4607 (2022).

\bibitem{seibert2011}
M.~M. Seibert, T.~Ekeberg, F.~R. N.~C. Maia, \emph{et~al.}, \enquote{Single mimivirus particles intercepted and imaged with an {{X-ray}} laser,} {\protect\JournalTitle{Nature}} \textbf{470}, 78--U86 (2011).

\bibitem{gorkhover2018}
T.~Gorkhover, A.~Ulmer, K.~Ferguson, \emph{et~al.}, \enquote{Femtosecond {{X-ray Fourier}} holography imaging of free-flying nanoparticles,} {\protect\JournalTitle{Nature Photonics}} \textbf{12}, 150--153 (2018).

\bibitem{hantke2014}
M.~F. Hantke, D.~Hasse, F.~R. N.~C. Maia, \emph{et~al.}, \enquote{High-throughput imaging of heterogeneous cell organelles with an {{X-ray}} laser,} {\protect\JournalTitle{Nature Photonics}} \textbf{8}, 943--949 (2014).

\bibitem{takayama2015}
Y.~Takayama, Y.~Inui, Y.~Sekiguchi, \emph{et~al.}, \enquote{Coherent {{X-Ray Diffraction Imaging}} of {{Chloroplasts}} from {{{\emph{Cyanidioschyzon}}}}{\emph{ merolae}} by {{Using X-Ray Free Electron Laser}},} {\protect\JournalTitle{Plant and Cell Physiology}} \textbf{56}, 1272--1286 (2015).

\bibitem{pan2022}
D.~Pan, J.~Fan, Z.~Nie, \emph{et~al.}, \enquote{Quantitative analysis of the effect of radiation on mitochondria structure using coherent diffraction imaging with a clustering algorithm,} {\protect\JournalTitle{IUCrJ}} \textbf{9}, 223--230 (2022).

\bibitem{kimura2014}
T.~Kimura, Y.~Joti, A.~Shibuya, \emph{et~al.}, \enquote{Imaging live cell in micro-liquid enclosure by {{X-ray}} laser diffraction,} {\protect\JournalTitle{Nature Communications}} \textbf{5}, 3052 (2014).

\bibitem{vanderschot2015}
G.~{van der Schot}, M.~Svenda, F.~R. N.~C. Maia, \emph{et~al.}, \enquote{Imaging single cells in a beam of live cyanobacteria with an {{X-ray}} laser.} {\protect\JournalTitle{Nature communications}} \textbf{6}, 5704 (2015).

\bibitem{kobayashi2021}
A.~Kobayashi, Y.~Takayama, T.~Hirakawa, \emph{et~al.}, \enquote{Common architectures in cyanobacteria {{Prochlorococcus}} cells visualized by {{X-ray}} diffraction imaging using {{X-ray}} free electron laser,} {\protect\JournalTitle{Scientific Reports}} \textbf{11}, 3877 (2021).

\bibitem{garcia-sucerquia2006}
J.~{Garcia-Sucerquia}, W.~Xu, S.~K. Jericho, \emph{et~al.}, \enquote{Digital in-line holographic microscopy,} {\protect\JournalTitle{Applied Optics}} \textbf{45}, 836 (2006).

\bibitem{gorniak2011}
T.~Gorniak, R.~Heine, A.~P. Mancuso, \emph{et~al.}, \enquote{X-ray holographic microscopy with zone plates applied to biological samples in the water window using 3rd harmonic radiation from the free-electron laser {{FLASH}},} {\protect\JournalTitle{Optics Express}} \textbf{19}, 11059 (2011).

\bibitem{abbey2008}
B.~Abbey, K.~A. Nugent, G.~J. Williams, \emph{et~al.}, \enquote{Keyhole coherent diffractive imaging,} {\protect\JournalTitle{Nature Physics}} \textbf{4}, 394--398 (2008).

\bibitem{williams2010}
G.~J. Williams, H.~M. Quiney, A.~G. Peele, and K.~A. Nugent, \enquote{Fresnel coherent diffractive imaging: {{Treatment}} and analysis of data,} {\protect\JournalTitle{New Journal of Physics}} \textbf{12}, 035020 (2010).

\bibitem{vassholz2021}
M.~Vassholz, H.~P. Hoeppe, J.~Hagemann, \emph{et~al.}, \enquote{Pump-probe {{X-ray}} holographic imaging of laser-induced cavitation bubbles with femtosecond {{FEL}} pulses,} {\protect\JournalTitle{Nature Communications}} \textbf{12}, 3468 (2021).

\bibitem{hagemann2021}
J.~Hagemann, M.~Vassholz, H.~Hoeppe, \emph{et~al.}, \enquote{Single-pulse phase-contrast imaging at free-electron lasers in the hard {{X-ray}} regime,} {\protect\JournalTitle{Journal of Synchrotron Radiation}} \textbf{28}, 52--63 (2021).

\bibitem{sandberg2014}
R.~L. Sandberg, C.~Bolme, K.~Ramos, \emph{et~al.}, \enquote{Ultrafast {{Imaging}} of {{Shocked Material Dynamics}} with {{X-ray Fee Electron Laser Pulses}},} in \emph{{{CLEO}}: 2014 {{Postdeadline Paper Digest}},}  (OSA, San Jose, California, 2014), p. STh5C.8.

\bibitem{schropp2015}
A.~Schropp, R.~Hoppe, V.~Meier, \emph{et~al.}, \enquote{Imaging {{Shock Waves}} in {{Diamond}} with {{Both High Temporal}} and {{Spatial Resolution}} at an {{XFEL}}.} {\protect\JournalTitle{Scientific reports}} \textbf{5}, 11089 (2015).

\bibitem{nagler2016}
B.~Nagler, A.~Schropp, E.~C. Galtier, \emph{et~al.}, \enquote{The phase-contrast imaging instrument at the matter in extreme conditions endstation at {{LCLS}},} {\protect\JournalTitle{Review of Scientific Instruments}} \textbf{87}, 103701 (2016).

\bibitem{seiboth2018}
F.~Seiboth, L.~B. Fletcher, D.~McGonegle, \emph{et~al.}, \enquote{Simultaneous 8.2 {{keV}} phase-contrast imaging and 24.6 {{keV X-ray}} diffraction from shock-compressed matter at the {{LCLS}},} {\protect\JournalTitle{Applied Physics Letters}} \textbf{112}, 221907 (2018).

\bibitem{vagovic2019}
P.~Vagovi{\v c}, T.~Sato, L.~Mike{\v s}, \emph{et~al.}, \enquote{Megahertz x-ray microscopy at x-ray free-electron laser and synchrotron sources,} {\protect\JournalTitle{Optica}} \textbf{6}, 1106 (2019).

\bibitem{mcnulty1992}
I.~McNulty, J.~Kirz, C.~Jacobsen, \emph{et~al.}, \enquote{High-{{Resolution Imaging}} by {{Fourier Transform X-ray Holography}},} {\protect\JournalTitle{Science}} \textbf{256}, 1009--1012 (1992).

\bibitem{eisebitt2004}
S.~Eisebitt, J.~L{\"u}ning, W.~F. Schlotter, \emph{et~al.}, \enquote{Lensless imaging of magnetic nanostructures by {{X-ray}} spectro-holography,} {\protect\JournalTitle{Nature}} \textbf{432}, 885--888 (2004).

\bibitem{marchesini2008}
S.~Marchesini, S.~Boutet, A.~E. Sakdinawat, \emph{et~al.}, \enquote{Massively parallel {{X-ray}} holography,} {\protect\JournalTitle{Nature Photonics}} \textbf{2}, 560--563 (2008).

\bibitem{guizar-sicairos2008b}
M.~{Guizar-Sicairos} and J.~R. Fienup, \enquote{Direct image reconstruction from a {{Fourier}} intensity pattern using {{HERALDO}},} {\protect\JournalTitle{Optics Letters}} \textbf{33}, 2668 (2008).

\bibitem{martin2014}
A.~V. Martin, A.~J. D'Alfonso, F.~Wang, \emph{et~al.}, \enquote{X-ray holography with a customizable reference,} {\protect\JournalTitle{Nature Communications}} \textbf{5}, 4661 (2014).

\bibitem{muller2013}
L.~M{\"u}ller, S.~Schleitzer, C.~Gutt, \emph{et~al.}, \enquote{Ultrafast {{Dynamics}} of {{Magnetic Domain Structures Probed}} by {{Coherent Free-Electron Laser Light}},} {\protect\JournalTitle{Synchrotron Radiation News}} \textbf{26}, 27--32 (2013).

\bibitem{willems2017}
F.~Willems, C.~{von Korff Schmising}, D.~Weder, \emph{et~al.}, \enquote{Multi-color imaging of magnetic {{Co}}/{{Pt}} heterostructures,} {\protect\JournalTitle{Structural Dynamics}} \textbf{4}, 014301 (2017).

\bibitem{vonkorffschmising2014}
C.~{von Korff Schmising}, B.~Pfau, M.~Schneider, \emph{et~al.}, \enquote{Imaging {{Ultrafast Demagnetization Dynamics}} after a {{Spatially Localized Optical Excitation}},} {\protect\JournalTitle{Physical Review Letters}} \textbf{112}, 217203 (2014).

\bibitem{egawa2024}
S.~Egawa, K.~Sakurai, Y.~Takeo, \emph{et~al.}, \enquote{Observation of mammalian living cells with femtosecond single pulse illumination generated by a soft {{X-ray}} free electron laser,} {\protect\JournalTitle{Optica}} \textbf{11}, 736 (2024).

\bibitem{levitan2020}
A.~L. Levitan, K.~Keskinbora, U.~T. Sanli, \emph{et~al.}, \enquote{Single-frame far-field diffractive imaging with randomized illumination,} {\protect\JournalTitle{Optics Express}} \textbf{28}, 37103 (2020).

\bibitem{sidorenko2016}
P.~Sidorenko and O.~Cohen, \enquote{Single-shot ptychography,} {\protect\JournalTitle{Optica}} \textbf{3}, 9--14 (2016).

\bibitem{kharitonov2022}
K.~Kharitonov, M.~Mehrjoo, M.~{Ruiz-Lopez}, \emph{et~al.}, \enquote{Single-shot ptychography at a soft {{X-ray}} free-electron laser,} {\protect\JournalTitle{Scientific Reports}} \textbf{12}, 14430 (2022).

\bibitem{pfeiffer2018}
F.~Pfeiffer, \enquote{X-ray ptychography,} {\protect\JournalTitle{Nature Photonics}} \textbf{12}, 9--17 (2018).

\bibitem{levitanCDTools2024}
A.~Levitan, M.~Cain, and A.~Kutakh, \enquote{{{CDTools}} 0.2.0 (python package),}  (2024). \url{https://github.com/cdtools-developers/cdtools}.

\bibitem{thibault2013}
P.~Thibault and A.~Menzel, \enquote{Reconstructing state mixtures from diffraction measurements,} {\protect\JournalTitle{Nature}} \textbf{494}, 68--71 (2013).

\bibitem{tripathi2014}
A.~Tripathi, I.~McNulty, and O.~G. Shpyrko, \enquote{Ptychographic overlap constraint errors and the limits of their numerical recovery using conjugate gradient descent methods,} {\protect\JournalTitle{Optics Express}} \textbf{22}, 1452 (2014).

\bibitem{marchesini2013}
S.~Marchesini, A.~Schirotzek, C.~Yang, \emph{et~al.}, \enquote{Augmented projections for ptychographic imaging,} {\protect\JournalTitle{Inverse Problems}} \textbf{29} (2013).

\bibitem{maiden2011}
A.~M. Maiden, M.~J. Humphry, F.~Zhang, and J.~M. Rodenburg, \enquote{Superresolution imaging via ptychography.} {\protect\JournalTitle{Journal of the Optical Society of America. A, Optics, image science, and vision}} \textbf{28}, 604--12 (2011).

\bibitem{levitan2026}
A.~Levitan, K.~Keskinbora, M.~Pancaldi, \emph{et~al.}, \enquote{{Replication Data for: Single shot imaging with randomized structured illumination at a free electron laser},}  (2026). \url{https://doi.org/10.7910/DVN/4LYYRS}.

\bibitem{levitan2022}
A.~Levitan and R.~Comin, \enquote{Error metrics for partially coherent wave fields,} {\protect\JournalTitle{Optics Letters}} \textbf{47}, 2322 (2022).

\bibitem{penczek2010}
P.~A. Penczek, \enquote{Resolution {{Measures}} in {{Molecular Electron Microscopy}},} in \emph{Methods in {{Enzymology}},}  vol. 482 (Elsevier, 2010), pp. 73--100.

\bibitem{vanheel2005}
M.~Van~Heel and M.~Schatz, \enquote{Fourier shell correlation threshold criteria,} {\protect\JournalTitle{Journal of Structural Biology}} \textbf{151}, 250--262 (2005).

\bibitem{vila-comamala2011}
J.~{Vila-Comamala}, A.~Diaz, M.~{Guizar-Sicairos}, \emph{et~al.}, \enquote{Characterization of high-resolution diffractive {{X-ray}} optics by ptychographic coherent diffractive imaging,} {\protect\JournalTitle{Optics Express}} \textbf{19}, 21333 (2011).

\bibitem{guizar-sicairos2012}
M.~{Guizar-Sicairos}, M.~Holler, A.~Diaz, \emph{et~al.}, \enquote{Role of the illumination spatial-frequency spectrum for ptychography,} {\protect\JournalTitle{Physical Review B - Condensed Matter and Materials Physics}} \textbf{86}, 100103(R) (2012).

\bibitem{donnelly2016}
C.~Donnelly, V.~Scagnoli, M.~{Guizar-Sicairos}, \emph{et~al.}, \enquote{High resolution hard x-ray magnetic imaging with dichroic ptychography,} {\protect\JournalTitle{Physical Review B}} \textbf{064421}, 1--5 (2016).

\bibitem{shapiro2020}
D.~A. Shapiro, S.~Babin, R.~S. Celestre, \emph{et~al.}, \enquote{An ultrahigh-resolution soft x-ray microscope for quantitative analysis of chemically heterogeneous nanomaterials,} {\protect\JournalTitle{Science Advances}} \textbf{6}, eabc4904 (2020).

\bibitem{fienup1997}
J.~R. Fienup, \enquote{Invariant error metrics for image reconstruction,} {\protect\JournalTitle{Applied Optics}} \textbf{36}, 8352 (1997).

\bibitem{lohmann1996}
A.~W. Lohmann, R.~G. Dorsch, D.~Mendlovic, \emph{et~al.}, \enquote{Space--bandwidth product of optical signals and systems,} {\protect\JournalTitle{Journal of the Optical Society of America A}} \textbf{13}, 470 (1996).

\bibitem{martin2012}
A.~Martin, N.~Loh, C.~Hampton, \emph{et~al.}, \enquote{Femtosecond dark-field imaging with an {{X-ray}} free electron laser,} {\protect\JournalTitle{Optics Express}} \textbf{20}, 13501 (2012).

\bibitem{capotondi2013}
F.~Capotondi, E.~Pedersoli, N.~Mahne, \emph{et~al.}, \enquote{Invited {{Article}}: {{Coherent}} imaging using seeded free-electron laser pulses with variable polarization: {{First}} results and research opportunities,} {\protect\JournalTitle{Review of Scientific Instruments}} \textbf{84}, 051301 (2013).

\bibitem{pfau2010}
B.~Pfau, C.~M. G{\"u}nther, S.~Schaffert, \emph{et~al.}, \enquote{Femtosecond pulse x-ray imaging with a large field of view,} {\protect\JournalTitle{New Journal of Physics}} \textbf{12}, 095006 (2010).

\end{thebibliography}

\end{document}

\maketitle

\section{Optic Design and Fabrication}\label{section:optic-design-fab}

\begin{figure}
\centerline{\includegraphics[width=170mm]{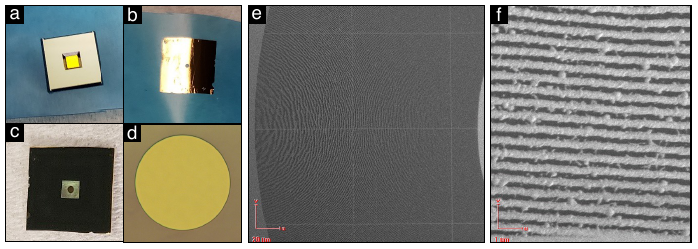}}
\caption[Fabrication of the \gls{rzp} optics.]{\textbf{Optics}. (a-d), The sketch-and-peel process for defining the beamstop. After ion beam lithography milling, the membrane is brought in contact with a low-tack tape (a), leaving a residual gold layer (b) on the tape, and the beamstop (c,d) on the membrane. (e-f), scanning electron microscope images of a large area (e) and the outermost zones (f) of the final ion-beam-lithography-fabricated \gls{rzp}.}\label{fig:optic_fab}
\end{figure}

The design file for the \gls{rzp} was produced using the method discussed in \cite{levitan2020}. The design parameters were: $\SI{200}{\nano\meter}$ outer zone width; $\SI{800}{\micro\meter}$ diameter; and a focal spot diameter of $\SI{40}{\micro\meter}$. An integrated beamstop was included with a diameter of $\SI{400}{\micro\meter}$. The designed optic has a focal distance of $\SI{7.743}{\milli\meter}$ at the design photon energy of $\SI{60}{\eV}$. The raster design file was generated at a $\SI{50}{\nano\meter}$ resolution, with computation performed on the MIT Supercloud cluster \cite{reuther2018}.

The optic was fabricated using a dedicated ion beam lithography platform (VELION, Raith, De) \cite{nadzeyka2012, keskinbora2013, keskinbora2018}. A $\SI{250}{\nano\meter}$ thick layer of gold for the beamstop was first deposited on a $\SI{100}{\nano\meter}$ thick silicon carbide membrane window using magnetron sputtering. The $\SI{400}{\micro\meter}$ diameter beamstop was then cut out using a sketch-and-peel process \cite{chen2016b}. A further $\SI{50}{\nano\meter}$ thick layer of gold was deposited on the chip for the \gls{rzp} pattern. The optic exposure pattern was divided into $10\times10$ $\SI{80}{\micro\meter}$ write fields. Each field was exposed using a $\SI{35}{\kilo\eV}$ beam of Au+ ions with a current of $\SI{344}{\pico\ampere}$ and a pixel size of $\SI{50}{\nano\meter}$.  The outer zone width of $\SI{200}{\nano\meter}$ was chosen to allow us to work at a low numerical aperture for this initial experiment. Images outlining the process can be found in figure \ref{fig:optic_fab}.

\section{FEL Data Collection}\label{section:data-collection}

Experiments were performed at the DiProI beamline of the FERMI \gls{fel} \cite{capotondi2013}. FEL-1 was used, producing pulses of light at $\SI{59.6}{\eV}$ with an estimated duration of $\SI{60}{\femto\second}$ \cite{allaria2012} and an estimated bandwidth $\frac{E}{\Delta E} > 1000$. The adjustable Kirkpatrick-Baez mirrors were underfocused, producing a beam of several millimeters in diameter at the plane of the optic. The optic was attached to an integrated mount with a matched order selecting aperture, and this mount was affixed to a set of x/y/z scanning motors and a goniometer for alignment with the optic axis.

To study the gold resolution test target, this spot was scanned over the test sample in a Fermat spiral pattern to generate a 300-shot ptychography dataset for calibration. Following that, 128 single-shot images were collected for analysis with \gls{rpi}. The diffraction was captured on a $2048\times2048$ CCD detector (Princeton Instruments MTE 2048B) with $\SI{13.5}{\micro\meter}$ pixels placed $\SI{135}{\milli\meter}$ downstream from the sample, run with $2\times2$ binning. Each single-shot diffraction pattern in the dataset used for single-shot \gls{rpi} reconstructions contained an average of $1.2\times10^7$ detected photons, equivalent to $490$ photons per $\SI{200}{\nano\meter}$ resolution element within the illuminated region. Accounting for absorption in the sample and the detector quantum efficiency, this corresponds to an estimated total pulse energy incident on the sample of $\SI{1.4}{\nano\joule}$, or a mean illumination fluence of $\SI{110}{\micro\joule\per\centi\meter\squared}$.

The imaging geometry, including incident fluence, was identical for the sample of Ta($\SI{3}{\nano\meter}$) / [Pt($\SI{3}{\nano\meter}$) / Co($\SI{1}{\nano\meter}$) / Ta($\SI{1.9}{\nano\meter}$)]x10 / Pt($\SI{2}{\nano\meter}$). Due to the weaker contrast and correspondingly lower resolution, a super-resolution model was not used for analysis, and instead the detector was cropped by 100 pixels to enable a better quality background subtraction to be performed. A ptychographic dataset in a Fermat spiral pattern was collected with 200 exposures for left-hand circular light, and with 215 exposures for right-handed circular light. Due to increased absorption in the magnetic sample, each single-shot diffraction pattern in the dataset used for single-shot \gls{rpi} reconstructions contained an average of only $2.8\times10^6$ detected photons, equivalent to $2.8\times10^3$ measured photons per $\si{\micro\meter}^2$.

\section{Resolution Estimation Methodology}\label{section:resolution}

We used the \gls{ssnr} to estimate the resolution of the ptychography and \gls{rpi} reconstructions reported in the paper. The \gls{ssnr} is a function of the \gls{frc} or \gls{fcr} and is used as an estimator of the signal-to-noise ratio within each of the radial bands defined in the calculation of the \gls{frc}.

\begin{figure}
\vspace*{-25mm}
\centerline{\includegraphics[width=150mm]{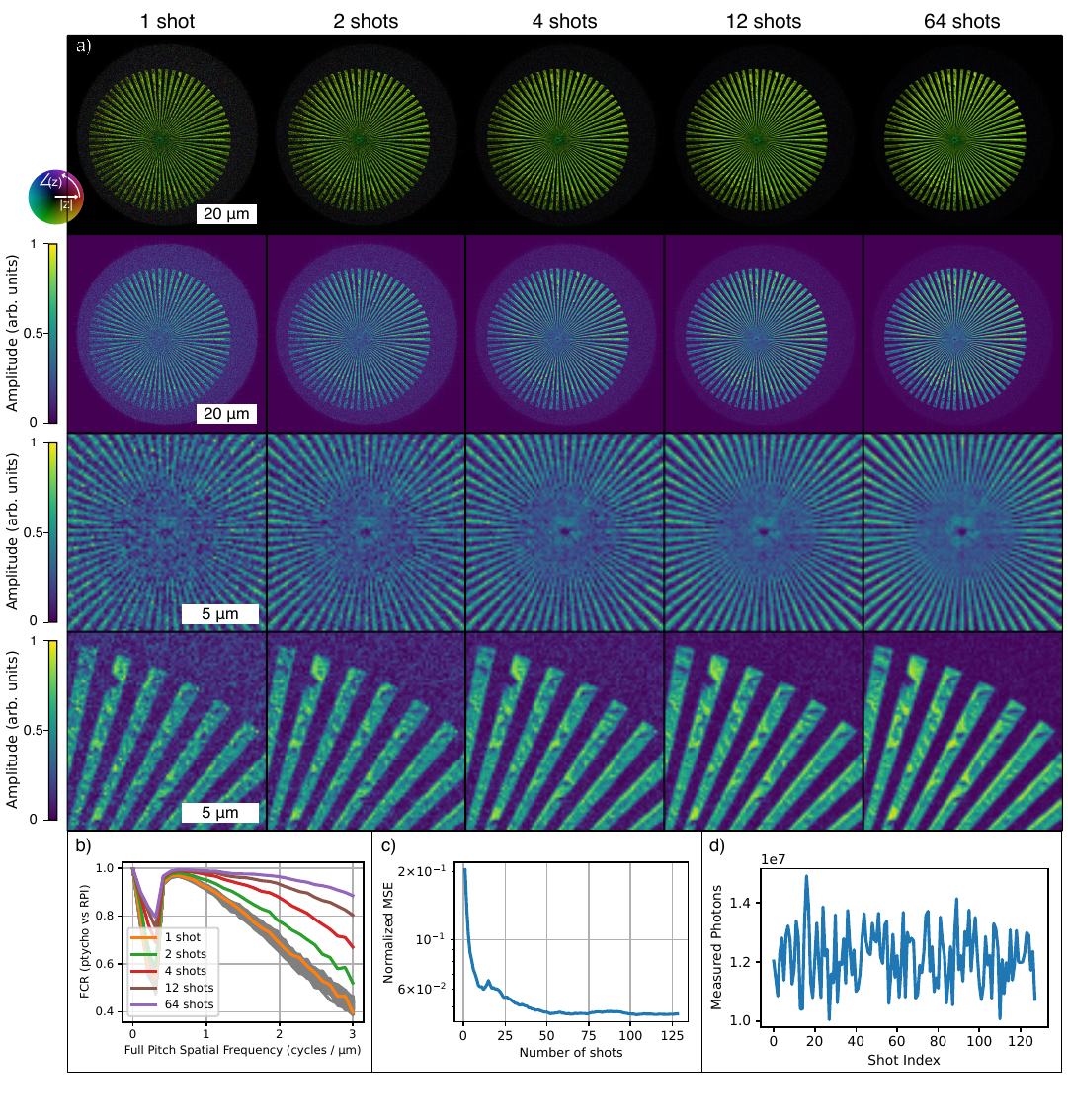}}
\caption[Further Details of Shot Number Dependence]{\textbf{Shot Number Dependence of RPI}. (a) The complex-valued image, amplitude, and two zoomed regions of the sample, as synthesized from a sum of 1, 2, 4, 12, and 64 shots. (b) The \gls{fcr} between each of the images above and the object, as reconstructed by ptychography. This was used to estimate the \gls{ssnr} reported in the main text (c) The dependence of the overall amplitude-minimized normalized mean squared error \cite{fienup1997} on the number of included shots, from 1 to 128.  (d) The number of photons measured on the detector for each of the 128 diffraction patterns.}\label{fig:shot_no_dep}
\end{figure}

In the classic setting for ptychography, two reconstructions are performed, each using half of the data. An appropriate estimator for the signal to noise ratio of the combined reconstruction, defined as a function of the \gls{frc} calculated between the two half-data reconstructions, is \cite{penczek2010, unser1987}:

\begin{equation}
\ssnr = 2 \frac{\frc}{1-\frc}
\end{equation}

The factor of 2 arises because we are estimating the \gls{ssnr} of the combined result.

However, we also want to be able to study noisy \gls{rpi} images by comparing single-shot reconstructions with much less noisy ptychography results that can be treated as a ground truth image. A common estimator for the \gls{ssnr} of the noisy image in this case uses the \gls{fcr}, \cite{penczek2010} and is:

\begin{equation}
\ssnr = \frac{\fcr^2}{1-\fcr^2}
\end{equation}

In our work, we compare the estimate of the \gls{ssnr} to the half-bit threshold $\text{SSNR} = 2^{0.5} - 1 \approx 0.414$, the threshold level proposed by Van Heel in the context of \gls{frc} curves \cite{vanheel2005} and most commonly used in the ptychography literature. It is more common to compare the \gls{frc} curve to one of the threshold curves defined by \cite{vanheel2005}. 

Here, we choose to use the \gls{ssnr} to enable comparison of the the ptychography results (analyzed via \gls{frc}) and \gls{rpi} results (analyzed via \gls{fcr}) on the same footing. To make it clear that this has not led to a reduction in the stringency of the resolution criteria, we have also included raw \gls{frc} and \gls{fcr} curves in Supplementary figures \ref{fig:ptycho} and \ref{fig:shot_no_dep}.

Finally, we estimated the number of measured photons per shot using the knowledge that the detector's output amplifier maps an input of 0-200,000 electrons to 16-bit output range, $0.32$ analog detector units per electron. We assumed that each detected photon produced one electron per every $\SI{3.6}{\eV}$ of photon energy. This leads to a final conversion of $5.4$ analog detector units per detected photon. 

\section{Linecut Methodology}\label{section:linecut}

We took a cut from the image along a circular arc that the spokes of the Siemens star intersect at a constant pitch. Because the path doesn't follow the grid axes, some form of interpolation is required. We used a sinc-interpolation, implemented by zero-padding in Fourier space. This is a natural choice because it matches the assumption baked into the \gls{rpi} model, which upsamples the low-resolution image with an identical zero-padding in Fourier space.

Once the image is sufficiently upsampled, the final linecut was extracted using a linear interpolation to sample explicit points defined along an arc. The center point of the arc's circle was identified manually.

\begin{figure}
\centerline{\includegraphics[width=150mm]{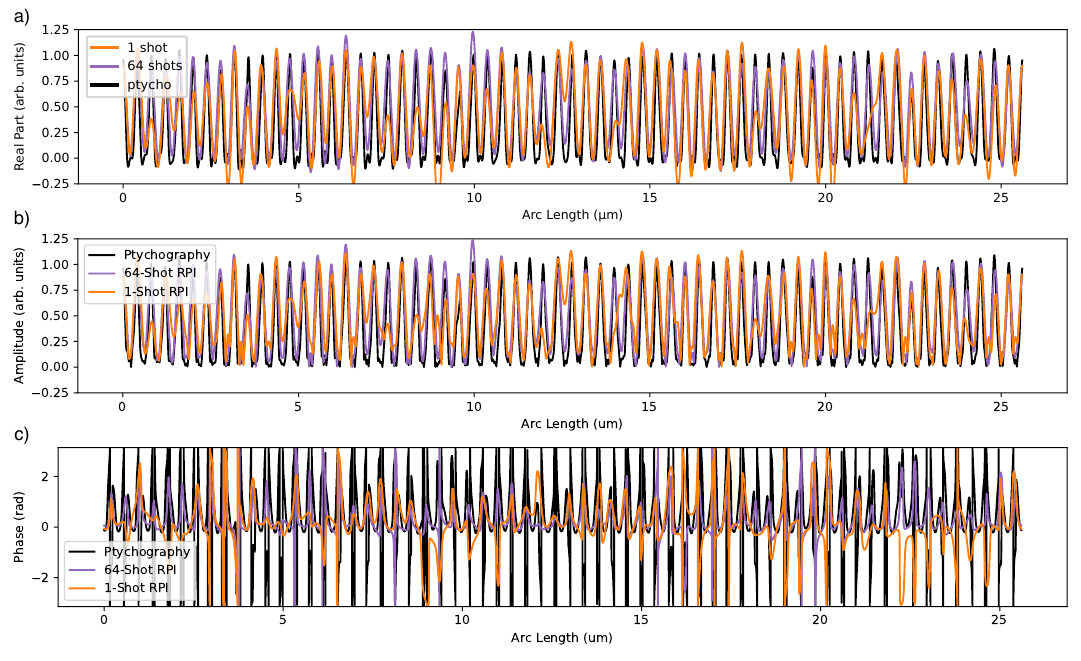}}
\caption[Full Arc Cut]{\textbf{Full Arc Cut}. (a) The real part, (b) the amplitude, and (c) the phase of a full cut from various reconstructions of the test sample, along an arc which intersects the spokes of the Siemens star at a $\SI{400}{\nano\meter}$ pitch. Included is a linecut from the ptychography reconstruction, from a representative \gls{rpi} reconstruction from one shot, and from an averaged \gls{rpi} reconstruction including 64 shots of data.} \label{fig:full_arc_cut}
\end{figure}

\section{Literature Search}\label{section:litsearch}

On November 13, 2023, we searched Clarivate web of science for the term:

\begin{quote}
(ALL=(``Free Electron Laser'') OR ALL=(``Free Electron Lasers'') OR ALL=(``FEL'') OR ALL=(``FELs'') OR ALL=(``XFEL'') OR ALL=(``XFELs'') OR ALL=(``Linac Coherent Light Source'') OR ALL=(``LCLS'') OR ALL=(``SACLA'') OR ALL=(``SPring-8 Angstrom Compact free electron LAser'') OR ALL=(``SwissFEL''))\\ AND\\ (ALL=(``Holography'') OR ALL=(``Holographic'') OR ALL=(``Ptychography'') OR ALL=(``Ptychographic'') OR ALL=(``Phase Retrieval'')  OR ALL=(``FXI'') OR ALL=(``CDI'') OR ALL=(``CMI') OR ALL=(``SSP'') OR ALL=(``Microscope'') OR ALL=(``Microscopes'') OR ALL=(``Microscopy'') OR ALL=(``Image'') OR ALL=(``Images'') OR ALL=(``Imaging''))
\end{quote}

The search term identifies papers which both mention \glspl{fel} and are plausibly related to quantitative phase microscopy. We included all generic terms for free electron laser of which we are aware, as well as the commonly used acronyms for every currently operating short wavelength free electron laser. We excluded the European XFEL and the Shanghai XFEL, as both names are already captured by the search term ``XFEL''. ``FERMI'' (Free Electron laser Radiation for Multidisciplinary Investigations) and ``FLASH'' (Free Electron LASer in Hamburg) were also excluded because they matched an unmanageably large number of papers unrelated to \glspl{fel}, and their full names are already captured by ``Free Electron Laser''. This search returned 5,094 results. 

Our goal was to identify all \gls{fel}-based publications that produced quantitative phase images using a method compatible with single-shot operation. Therefore, we excluded papers that only reported ptychography results, e.g. \cite{pound2020, kharitonov2021}. We did include Fourier transform holography results even when a sum over multiple shots was used (e.g. \cite{vonkorffschmising2014} or \cite{willems2017}). We also included in-line holography publications where phase retrieval was attempted, but excluded publications where raw phase-contrast images were reported without any quantitative inversion of the full 2D image (e.g. \cite{beckwith2017, buakor2022, faenov2018, maeda2020, makarov2023, nagler2015, nagler2016, sandberg2014, sawada2023, schropp2012, seiboth2018}).

To identify the relevant papers, we started by reading the title. For all papers which weren't obviously false positives (e.g. papers about the cat-allergy-related FEL protein), we read the abstract and viewed the figures, reading the full paper only if either the abstract or figures indicated the presence of data meeting our conditions. In cases where multiple papers used the same dataset, we only included the publication that reported the highest-quality images. In the end, we found 96 papers which reported unique quantitative phase contrast images captured by \glspl{fel} \cite{bajt2008, barty2008, bielecki2019, bogan2008, bogan2010, bogan2010a, boutet2008, branden2019, capotondi2013, chapman2006, clark2015a, daurer2017, diao2020, fan2016, fan2022, feinberg2022, gallagher-jones2014, gao2023, gorkhover2018, gorniak2011, gunther2011, hagemann2021, hagström2022, hantke2014, hau-riege2010, hodge2022, huang2018b, huang2020a, ihm2019, johnson2022, jones2016, jung2020, jung2021, kameda2017, kang2020b, kassemeyer2012, kharitonov2022, kimura2014, kimura2020, kobayashi2014, kobayashi2021a, lee2021, loh2012, mancuso2009, mancuso2010, marchesini2008, martin2011, martin2012, martin2012a, martin2014, matsumoto2022, nakasako2013, nam2016, nishino2010, nishiyama2020, oroguchi2015, oroguchi2018, osterhoff2021, pan2022, park2013, pedersoli2013, pfau2010, rosenhahn2009, sandberg2014, schropp2015, seibert2010, seibert2011, sekiguchi2014, sekiguchi2016, sekiguchi2017, shin2023, song2014, sun2018a, sung2021, suzuki2020, suzuki2022a, takahashi2013a, takayama2015, takayama2016, tanyag2015, ulmer2023, vagovic2019, vanderschot2015, vassholz2021, vassholz2023, vonkorffschmising2014, wang2012a, weder2017, wei2016, willems2017, xian2022, yang2019a, yoon2014, yoshida2015, yumoto2022, zhuang2022}.

Within each paper, we identified the image with the highest resolution and space-bandwidth product, adding multiple results if these images were not the same or if images captured with multiple methods were presented. This resulted in 101 unique images. Finally, we added one additional phased dataset \cite{muller2013} which had not been indexed in web of science because it was published in Synchrotron Radiation News. This brought the total to 97 papers and 102 images.

From each image, we estimated the full-pitch resolution, the field of view, and the space-bandwidth product. The resolution was calculated based on the author's claims, if possible. We accepted all author resolution claims based on image features - for example, 10\%-90\% rise, \gls{frc} or phase retrieval transfer function calculations, or visibility of lines in a test target. If a half-pitch resolution was reported, we converted it to full-pitch. In cases where the pixel pitch was reported as the resolution, where the resolution was derived from a numerical aperture-based argument, or no resolution was reported, we attempted to estimate the image resolution using the provided images. If the images appeared close to pixel-resolved, we simply used the pixel pitch. Otherwise, we manually estimated the length scale of the smallest features visible in the images. Because of the variety of methods used, we suggest that the resolution estimates should only be trusted to within roughly a factor of 2.

The field of view was estimated as the area of a box surrounding the imaged region. In cases where the field of view was densely populated, the space-bandwidth product was calculated using this field of view as the imaged area. In cases where the object had a narrow, sparse, or disjoint support, the actual imaged area can be an order of magnitude or more less than the area contained within this bounding box. In these cases we attempted to estimate the area of the imaged region. For cases where the object has disjoint support, the area of each disjoint region was separately estimated and then added. For circular supports, the area of the circle was calculated. Finally, in cases where a finite support constraint was used and the number of pixels in that support region was reported, the number of pixels was used to calculate the area of the field of view.

Finally, the space-bandwidth product was calculated as follows:

\begin{equation}
\text{SBP} = A_\text{real} A_\text{Fourier} = \pi A_\text{real}{\text{res}^2}.
\end{equation}

Where $\text{res}$ is expressed in cycles per length. A spreadsheet containing all the papers included in the search, as well as the resolution and field of view estimates, is available with the replication data and code at reference \cite{levitan2026}.

\section{Additional Figures}

\begin{figure}
\centerline{\includegraphics[width=150mm]{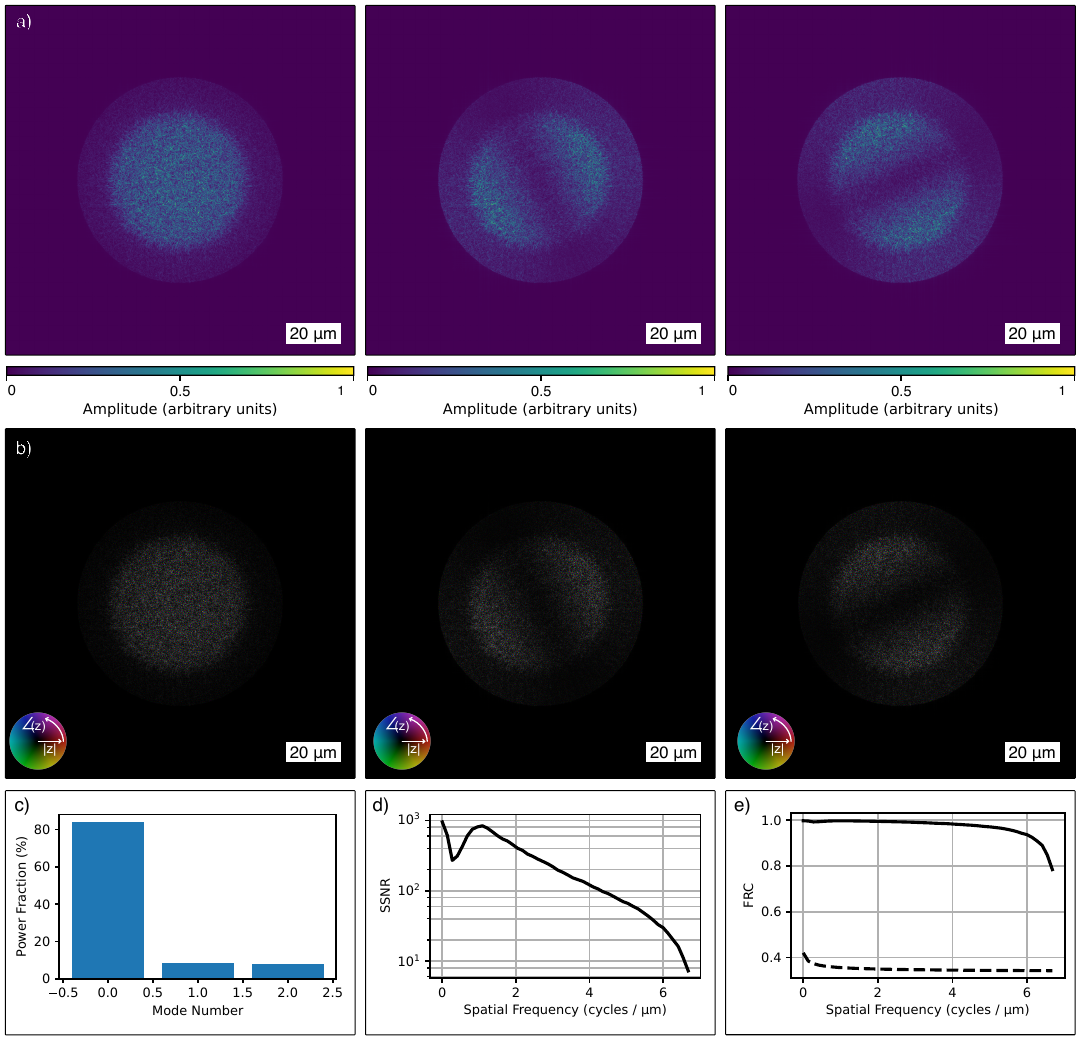}}
\caption[Further Details of Ptychography Calibration]{\textbf{Ptychography Calibration Details}. (a) The amplitudes of the three reconstructed probe modes, normalized to the brightest pixel within each image. (b) The complex-valued probe modes. (c) The power ratio of the three modes. (d) The \gls{ssnr} of the reconstructed object, as estimated via the \gls{frc}. (e) The \gls{frc} of the reconstructed object, compared with a half-bit threshold.}\label{fig:ptycho}
\end{figure}

\begin{figure}
\vspace*{-30mm}
\centerline{\includegraphics[width=150mm]{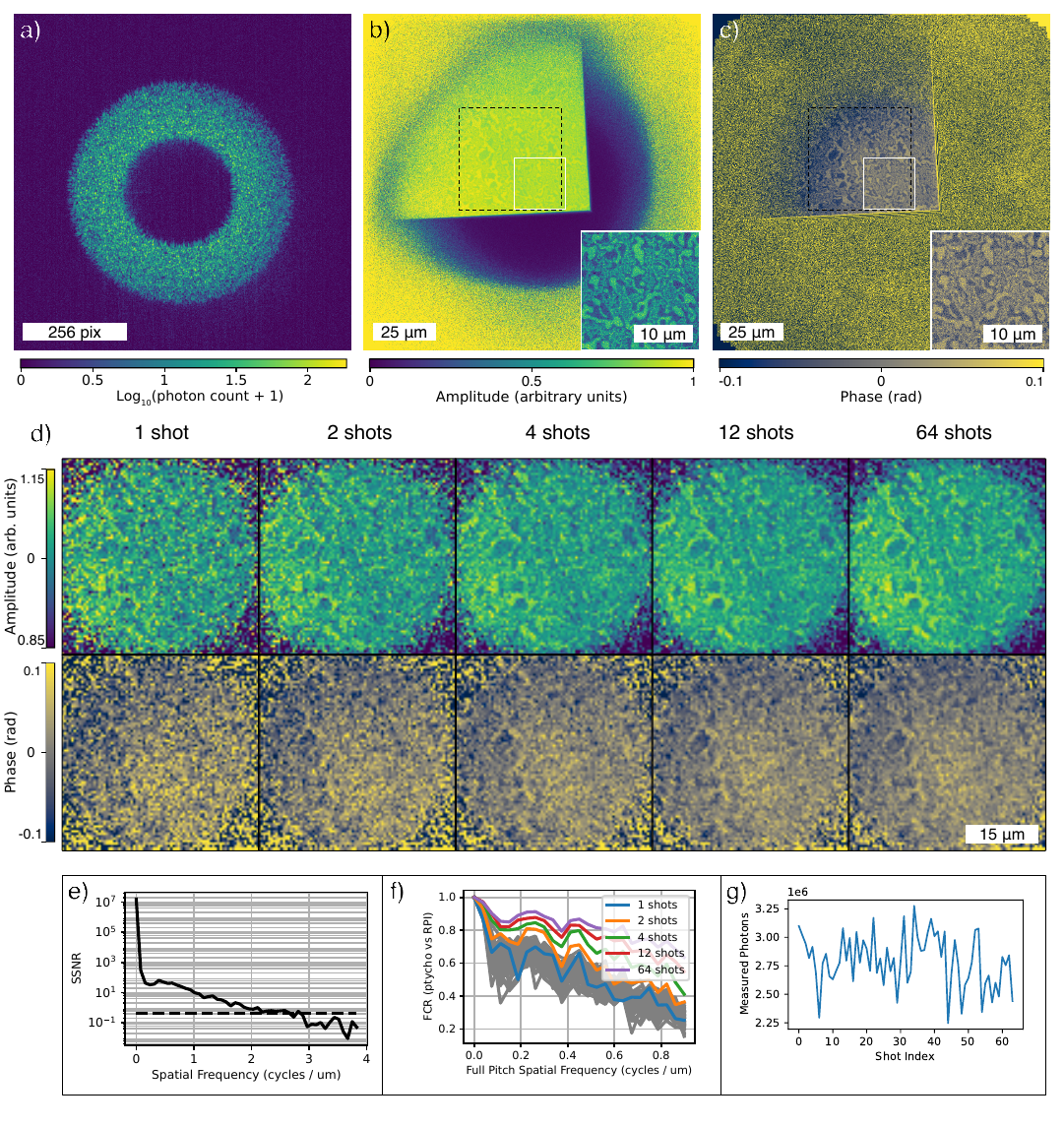}}
\caption[Further information about the magnetic imaging]{\textbf{Magnetic Imaging Details}. (a) A representative detector image after background subtraction, used for the reported single-shot RPI reconstruction. Ptychographic amplitude (b) and phase (c) maps using left-hand circularly polarized light. The region which overlaps with the \gls{rpi} reconstructions is shown in a dashed black line, and the region used for \gls{frc} calculation is outlined in white (inset). The color scale of the overview amplitude image in (b) is set to preserve the zero level, and the color scale of the inset matches that in Figure 3b from the main manuscript. (d) Shot-number dependence of the averaged \gls{rpi} images. (e) \gls{ssnr}, calculated via \gls{frc}, of the magnetic ptychography images. The threshold is set at 0.414 as discussed in Section \ref{section:resolution}. (f) \gls{fcr} curves, calculated against the ptychography reconstruction, for each image in (d). (g) The shot-to-shot intensity of the data used for magnetic \gls{rpi}.}\label{fig:magnetic}
\end{figure}

\begin{figure}
\vspace*{-20mm}
\centerline{\includegraphics[width=150mm]{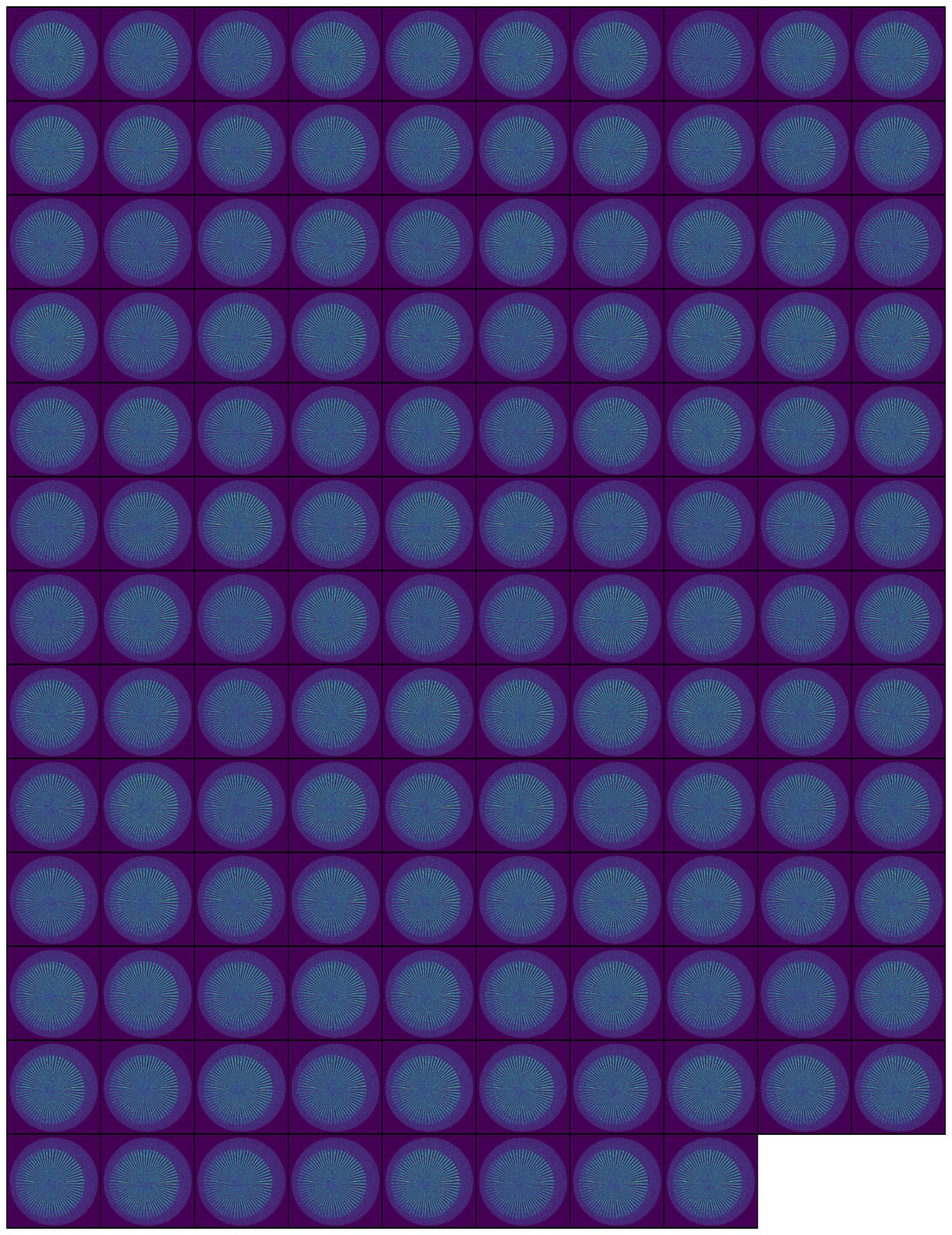}}
\caption[All \gls{rpi} reconstructions.]{\textbf{All \gls{rpi} Reconstructions}. The amplitude of all 128 single-shot \gls{rpi} reconstructions.}\label{fig:all_rpi}
\end{figure}

\begin{figure}
\vspace*{-20mm}
\centerline{\includegraphics[width=150mm]{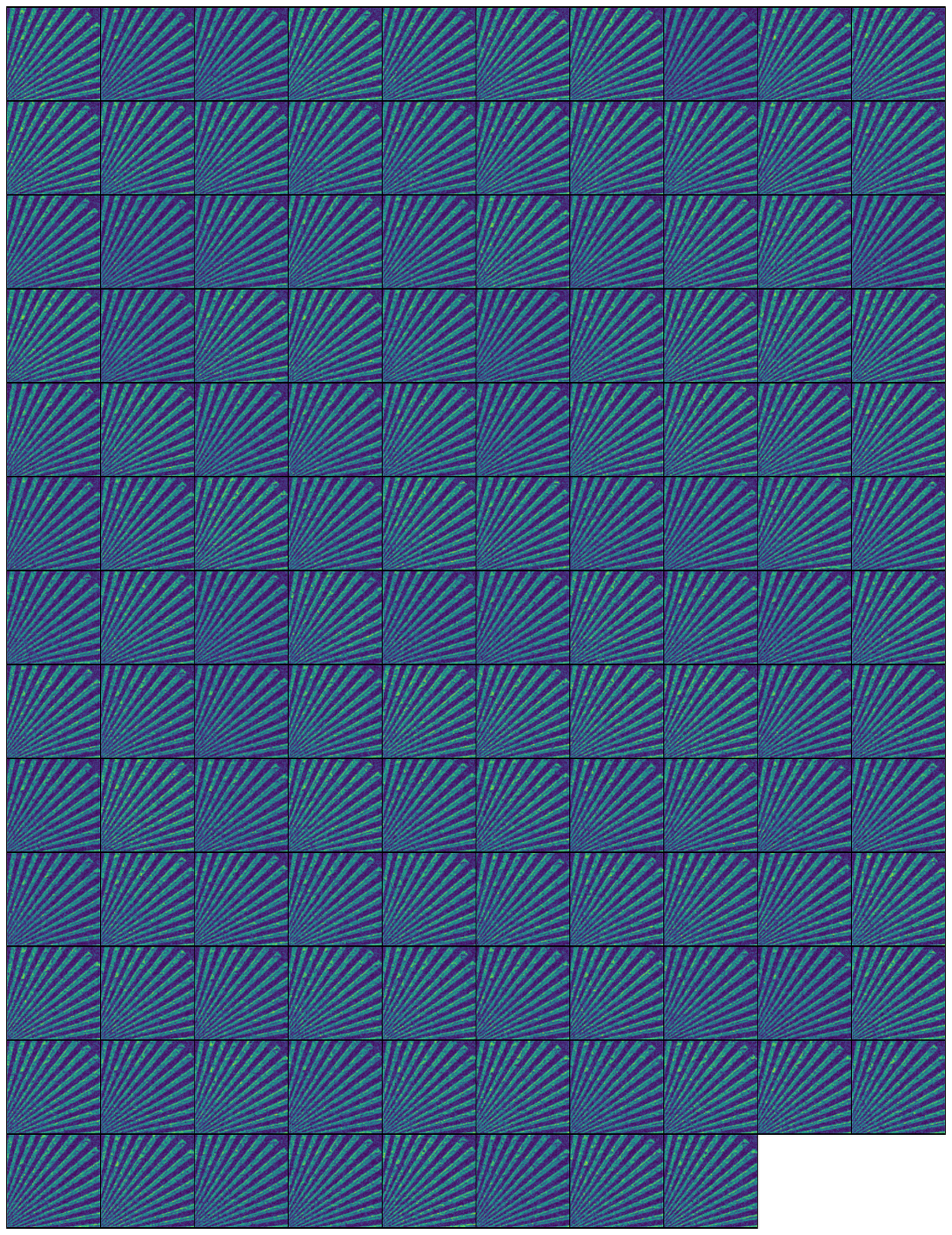}}
\caption[All \gls{rpi} reconstructions.]{\textbf{Cropped \gls{rpi} Reconstructions}. Closeup view of the amplitude of all 128 single-shot \gls{rpi} reconstructions.}\label{fig:all_rpi_zoom}
\end{figure}

\FloatBarrier
\bibliography{main}